\documentclass[10pt,aps,prb,superscriptaddress,nobibnotes,notitlepage,twocolumn]{revtex4-2}

\usepackage{mathptmx}
\usepackage{mathtools}
\usepackage[normalem]{ulem}
\usepackage{graphicx}
\graphicspath{{./Figs}}

\usepackage{hyperref}
\hypersetup{colorlinks=true, linkcolor=blue, citecolor=blue, urlcolor=blue}

\newcommand{\beq}[1]{\begin{equation}\label{#1}}
\newcommand{\eeq}{\end{equation}}
\newcommand{\eep}{\;.\end{equation}}
\newcommand{\eec}{\;,\end{equation}}

\newcommand{\lb}{\left(}
\newcommand{\rb}{\right)}

\newcommand{\la}{\lambda}
\renewcommand{\th}{\theta}

\newcommand{\D}{\Delta}
\newcommand{\G}{\Gamma}

\DeclareMathAlphabet{\mathcal}{OMS}{cmsy}{m}{n}

\newcommand{\M}{\mathcal{M}}
\newcommand{\N}{\mathcal{N}}

\newcommand{\bvec}[1]{\mathbf{#1}}

\newcommand{\kv}{\bvec{k}}

\newcommand{\rv}{\bvec{r}}

\newcommand{\Gfp}{\Gamma_{0+}}
\newcommand{\Gfm}{\Gamma_{0-}}
\newcommand{\Gfpm}{\Gamma_{0\pm}}
\newcommand{\Gdp}{\Gamma_{1+}}
\newcommand{\Gdm}{\Gamma_{1-}}

\newcommand{\Mdp}{\mathrm{M}_{1+}}

\newcommand{\HarvardQSE}{Quantum Science and Engineering, Harvard University, Cambridge, Massachusetts 02138, USA}
\newcommand{\HarvardPhysics}{Department of Physics, Harvard University, Cambridge, Massachusetts 02138, USA}
\newcommand{\HarvardSeas}{John A.~Paulson School of Engineering and Applied Sciences, Harvard University, Cambridge, Massachusetts 02138, USA}
\newcommand{\FSU}{Department of Physics, Florida State University, Tallahassee, Florida 32306, USA}
\newcommand{\NTU}{School of Electrical and Electronic Engineering, Nanyang Technological University Singapore, 50 Nanyang Avenue, 639798, Singapore}

\begin{document}

\title{Wavefunction textures in twisted bilayer graphene from first principles}

\author{Albert Zhu}
\affiliation{\HarvardQSE}

\author{Daniel Bennett}
\email{daniel.bennett@ntu.edu.sg}
\affiliation{\HarvardSeas}
\affiliation{\NTU}

\author{Daniel T.~Larson}
\affiliation{\HarvardPhysics}

\author{Mohammed M. Al Ezzi}
\affiliation{\HarvardSeas}

\author{Efstratios Manousakis}
\affiliation{\FSU}

\author{Efthimios Kaxiras}
\affiliation{\HarvardSeas}
\affiliation{\HarvardPhysics}

\begin{abstract}
Motivated by recent experiments probing the wavefunctions of magic-angle twisted bilayer graphene (tBLG), we perform large-scale first-principles calculations of tBLG with full atomic relaxation across a wide range of twist angles down to $0.99^\circ$. 
Focusing on the magic angle, we compute wavefunctions of the low energy bands, resolving atomic-scale details and moir\'e-scale patterns that form triangular, honeycomb, and Kagome lattices.
By tuning the interlayer interactions, we illustrate the formation of the flat bands from isolated monolayers and the emergence of the band inversion and fragile topology at a sufficiently large interaction strength.
We identify strong indicators of a new phase transition with increasing interlayer interaction strength, achievable with external pressure or a decrease in the twist angle.
When this transition occurs, the upper and lower flat bands exchange their wavefunction character and symmetry eigenvalues, which may be correlated with the appearance of superconductivity with electron doping below the magic angle.
Our study demonstrates the feasibility of using first-principles wavefunctions to help interpret experimental signatures of topological and correlated phases in tBLG.
\end{abstract}

\maketitle


\section{Introduction}
Twisted bilayer graphene (tBLG) near the magic angle has emerged as a prominent platform for studying superconductivity \cite{cao2018unconventional,lu2019superconductors,yankowitz2019tuning,oh2021evidence,tanaka2025superfluid,banerjee2025superfluid}, correlated insulator behavior \cite{cao2018correlated,codecido2019correlated,nuckolls2023quantum,tian2024dominant,krishna2025terahertz}, and topological phenomena such as
the anomalous Hall effect \cite{sharpe2019emergent,serlin2020intrinsic}, Chern insulators \cite{nuckolls2020strongly,wu2021chern,choi2021correlation,stepanov2021competing}, and fractional Chern insulators \cite{xie2021fractional}.
With recent experimental advances, it is now possible to measure and spatially resolve fundamental properties of tBLG with high precision, providing insight into these exotic phases.
For example, quantum twisting microscopy allows direct measurement of electronic bands \cite{inbar2023quantum} and phonons \cite{birkbeck2025quantum}.
With scanning tunneling microscopy (STM) it is possible to probe the local density of states with high spatial and energetic resolution, providing microscopic detail of the electronic wavefunctions at different energies \cite{nuckolls2023quantum}.
These measurements have revealed novel orders such as Kekulé spirals and vortex textures \cite{kwan2021kekule}, which has motivated the need for accurate theoretical modeling, both at the atomic and moiré scales.

A variety of theoretical methods have been developed to study the electronic structure of magic-angle tBLG \cite{carr2020electronic}, ranging from early continuum models \cite{dos2007graphene,Bistritzer2011,mele2011band,lopes2012continuum} to more realistic theories, which include \textit{ab initio} tight-binding models \cite{trambly2010localization,jung2014accurate,fang2016electronic,carr2017twistronics}, ``exact" $\kv\cdot {\bf p}$ continuum models \cite{carr2019exact,fang2019angle,guinea2019continuum,kang2023pseudomagnetic,vafek2023continuum,miao2023truncated}, and minimal phenomenological models that use the smallest number of bands to capture the low-energy physics of tBLG \cite{zou2018band,po2018origin,koshino2018maximally,kang2018symmetry,po2019faithful,tarnopolsky2019origin,carr2019exact,carr2019derivation,bernevig2021twisted,song2022magic,bennett2024twisted}.
While these approaches provide valuable insight, they rarely capture microscopic details, such as descriptions of the wavefunctions throughout 3D space, necessary for describing the fine features of the wavefunctions observed in STM experiments \cite{nuckolls2023quantum}.
First-principles density functional theory (DFT) can resolve these details, but its high computational cost has limited previous studies to large twist angles or specific properties such as atomic relaxation and electronic band structures \cite{uchida2014atomic,lucignano2019crucial,yananose2021chirality,song2019all}.
However, the use of an optimized local basis \cite{bennett2025accurate} has allowed us to significantly expand the capabilities of DFT as a tool for studying the wavefunctions of tBLG.

Here, we perform comprehensive DFT calculations of tBLG spanning a wide range of twist angles, from $21.79^\circ$ down to $0.99^\circ$, with supercells containing up to 13,468 atoms, and present results focusing on the microscopic and moiré-scale structure of the low-energy wavefunctions near the magic angle.
Our calculations reveal detailed microscopic and moiré-scale textures in the wavefunctions that can be connected with real-space experimental observations. 
On the moiré scale, we observe that the wavefunction amplitudes concentrate selectively on the AA, AB/BA, and domain wall (DW) regions, giving rise to distinctive triangular, honeycomb, and Kagome lattices, respectively.

We also examine how interlayer interactions influence the electronic properties of tBLG, demonstrating the onset of band inversion and fragile topology \cite{zou2018band,po2019faithful,ahn2019failure,song2019all}.
Notably, we identify the existence of a critical minimum interlayer coupling strength required for the emergence of the band inversion. 
Furthermore, we discover strong indication of a new
phase transition occurring at a critical pressure (or twist angle) in which the flat bands touch each other at $\G$, resulting in an exchange of wavefunction character between the upper and lower flat bands. 
This finding can guide experimental measurements of topological phases, and overall, our calculations provide the groundwork for future atomic-scale computational studies of correlated states in tBLG.

\begin{figure}[t!]
\centering
\includegraphics[width=1.0\columnwidth]{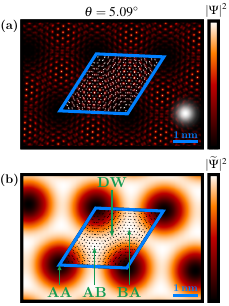}
\caption{%
Wavefunction of tBLG in the lower flat band at $\Gamma$ for a twist angle of $\theta=5.09^{\circ}$.
{\bf (a)} Modulus squared of the atomic-scale wavefunction $\vert\Psi\vert^2$, obtained from first-principles calculations.
{\bf (b)} Moiré-scale  probability density $\vert \widetilde{\Psi} \vert^2 = \vert\Psi\vert^2 * \N$, obtained by a convolution with a Gaussian $\N$ with a standard deviation of $2.5$ \AA, sketched in white above the $1$ nm scale bar in (a).
The atomic positions are denoted by white and black dots in (a) and (b), respectively. 
The different stacking regions, AA, AB, BA and domain walls (DW), are labeled in (b).
}
\label{Fig1}
\end{figure}

\begin{figure*}[t]
\centering
\includegraphics[width=\textwidth]{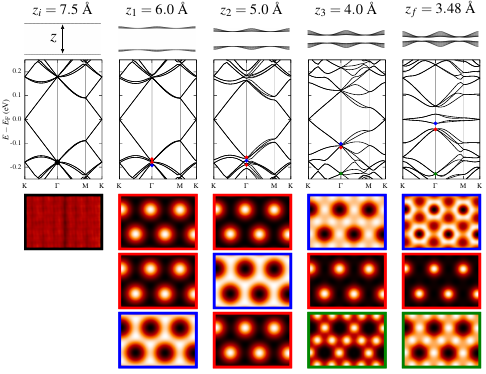}
\caption{%
Emergence of the band inversion in magic-angle tBLG ($\th = 1.08^{\circ}$). 
Each column shows a schematic atomic structure, the band structure, and wavefunction probability densities of the flat bands and dispersive bands for select values of the mean bilayer separation $z$, which effectively dictates the interlayer interaction strength. 
The band inversion is demonstrated by the energy of the AA$_z$ state and its precursors at larger $z$ (indicated by blue points in the band structure and blue borders around the probability densities) relative to the energy of the AA state(s) (indicated in red) in each column. 
DW states are indicated in green.
}
\label{Fig2}
\end{figure*}

\begin{figure*}[t]
\centering
\includegraphics[width=\textwidth]{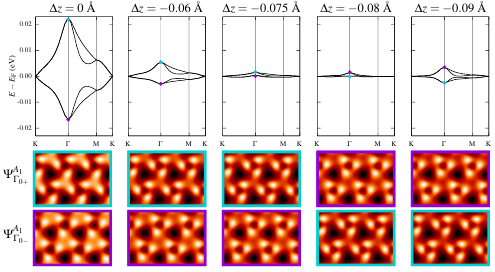}
\caption{%
Electronic band structure and wavefunctions at $\Gfpm$ of magic-angle tBLG as a function of $\D z = z-z_f$.
The probability density of the projection of the wavefunction onto layer 1 and sublattice $A$, $\Psi^{A_1}_{\Gfpm}$, is shown.
Cyan and purple borders (and the corresponding points on the band structures) denote wavefunction projections with different orientations.
}
\label{Fig3}
\end{figure*}


\section{Results}
\subsection{Microscopic and Moir\'e-scale Wavefunctions}
DFT calculations were performed using the {\sc siesta} code \cite{soler2002siesta,SM}, which employs a local atomic orbital basis set recently optimized for graphene \cite{junquera2001numerical,bennett2025accurate}.
After relaxing the atomic structures, the electronic band structures were obtained, as well as the wavefunctions of the flat bands and the neighboring dispersive bands at the $\G$, K and M high-symmetry points of the moir\'e Brillouin zone (mBZ).
A timing analysis is shown in Fig.~S6 \cite{SM}.
Fig.~\ref{Fig1} shows the $\G$-point wavefunction of the lower flat band for tBLG with $\th=5.09^{\circ}$.
The microscopic wavefunction probability density $\vert\Psi\vert^2$ (Fig.~\ref{Fig1} (a)) shows an intricate pattern of bright spots in the AB and BA regions that meet and form interference patterns in the DWs.
To visualize the moir\'e-scale behavior of the wavefunctions, we perform a convolution of $\vert\Psi\vert^2$ with a 2D Gaussian \cite{SM}.
The resulting moir\'e-scale wavefunction probability density $\vert\widetilde{\Psi}\vert^2$ exhibits smooth features coinciding with different regions of the moir\'e pattern, such as a honeycomb lattice shown in Fig.~\ref{Fig1} (b) arising from high wavefunction density in the AB and BA domains and low wavefunction density at the AA sites.
(All following figures, except Fig.~\ref{Fig4}, show the moir\'e-scale probability density $\vert\widetilde{\Psi}\vert^2$).

The localization of the wavefunctions depends on both the band and $\mathbf{k}$-point.
To succinctly identify particular wavefunctions, we introduce the following notation: X$_{n\pm}$, where $\mathrm{X}=\G,\ \mathrm{K},\ \mathrm{M}$ represents the high-symmetry $\kv$-points of the mBZ, and $n = 0,1,2,\ \ldots$ is the band index, with 0 referring to the flat bands, 1 referring to the first set of dispersive bands, and so on.
The sign indicates whether the bands are above ($+$) or below ($-$) the charge neutrality point.
For example, $\Gfm$ is the lower flat band at $\G$, and $\Mdp$ is the first dispersive conduction band at M.
To describe the localization of the wavefunctions, we follow the notation of Ref.~\cite{carr2019exact} and highlight four categories:
(i) ``AA" wavefunctions that are localized at AA sites and form a triangular lattice, 
(ii) ``AA$_z$'' wavefunctions that form rings around AA sites, 
(iii) ``DW'' wavefunctions that have weight along domain walls and form a Kagome lattice, and 
(iv) ``AB/BA" wavefunctions that are centered in AB/BA domains and form a honeycomb lattice. In Ref.~\cite{carr2019exact} the notation ``AA$_z$" refers to a moir\'{e}-scale orbital with $p_z$-like character that has circular symmetry but is odd under $z\rightarrow -z$. Here we use ``AA$_z$" to focus mainly on the in-plane distribution of the wavefunction that has a characteristic ring shape centered on the AA sites.
All wavefunction probability density plots have color-coded borders corresponding to these four categories, respectively red, blue, green, and pink.
Examples of all four are shown in Fig.~S1 \cite{SM}. 
Although it is not the focus of this work, we show the evolution of select wavefunctions as a function of twist angle in Fig.~S2 \cite{SM}, with a more detailed analysis to be done in future work.

\subsection{Emergence of Band Topology}

The wavefunctions of tBLG provide a useful framework to investigate its electronic and topological properties.
Although the concepts of fragile topology and band inversion in tBLG are well-established, their physical origin has not been explored; 
thus, we use the \textit{ab initio} wavefunctions to illustrate how these topological aspects arise as a function of the interlayer interactions.
While it is not possible to directly tune the strength of these interactions in DFT calculations, we can instead tune the interlayer spacing. 
This modifies the vdW interactions between the layers, allowing us to systematically investigate their effect on the electronic properties of tBLG.
Fig.~\ref{Fig2} shows the evolution of the band structure and $\G$-point wavefunctions of magic-angle tBLG ($\theta=1.08^\circ$) as we systematically interpolate between two flat isolated graphene monolayers and relaxed tBLG. 
We define the position of an atom in an interpolated structure, $\rv_n$, as a linear interpolation between the corresponding atomic positions in the isolated graphene monolayers, $\rv_i$, and the fully relaxed structure, $\rv_f$:
\beq{eq:interpolation}
\rv_{n} = \lb 1 - \frac{z_{n} - z_{f}}{z_{i} - z_{f}}\rb\rv_{f} + 
\lb \frac{z_{n} - z_{f}}{z_{i} - z_{f}}\rb\rv_{i}
\eec
where $z_n$ is the mean layer separation of the $n^{\mathrm{th}}$ interpolated structure, $z_f$ is the mean layer separation of the fully relaxed structure, and $z_i=7.5~\mathrm{\AA}$ ensures negligible vdW interactions between the flat graphene monolayers.
The atomic coordinates $\rv_n$ remain fixed during calculations, although we expect the results to qualitatively be unaffected by this construction.

\begin{figure*}[t]
\centering
\includegraphics[width=1.6\columnwidth]{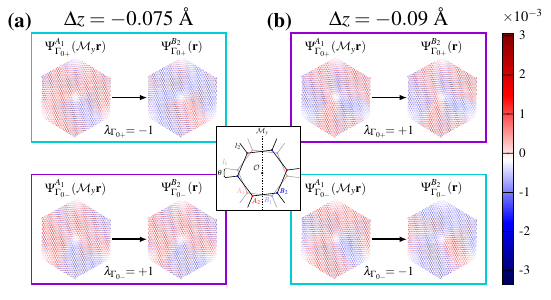}
\caption{%
Transformation of the sublattice-projected microscopic wavefunctions of magic-angle tBLG under the $\M_y$ symmetry (shown in the center).
{\bf (a)} Layer and sublattice projections of the flat band microscopic wavefunctions $\Psi^{A_1}_{\Gfpm}$ and $\Psi^{B_2}_{\Gfpm}$ for $\Delta z = -0.075~\mathrm{\AA}$ with $\M_y$ applied to $\Psi^{A_1}_{\Gfpm}$.
The inverted colors of the microscopic wavefunctions show that the upper flat band has a $\M_y$ eigenvalue of $\la_{\Gfp}\! = -1$.
{\bf (b)} The same as (a), but for $\Delta z = -0.09$ \AA.
In this case, $\la_{\Gfm}\! = -1$.
}
\label{Fig4}
\end{figure*}

When the graphene layers are isolated, the bands fold into the mBZ, resulting in highly degenerate, delocalized wavefunctions at $\G$ (Fig.~\ref{Fig2}, first column).
When the layers move closer together and begin to interact, as shown in the second and third columns of Fig.~\ref{Fig2}, this degeneracy is lifted and two AA wavefunctions emerge, 
as well as a precursor to the AA$_z$ wavefunction.
Decreasing the separation further, the precursor to the AA$_z$ wavefunction shifts up in energy, eventually joining the flat band manifold (fourth column) and localizing in rings around the AA sites in the fully-relaxed structure (last column). 
Additionally, when the layers are close together, a DW state becomes prominent in the second dispersive band (fourth and last columns).

Band inversion and the opening of mini-gaps between the flat and dispersive bands is a defining feature of the fragile topology of tBLG \cite{zou2018band,po2019faithful,ahn2019failure,song2019all}.
We find that a minimum interlayer interaction strength is necessary to lift wavefunction degeneracies and produce the characteristic topological band structure of tBLG.
The onset of this band topology, and its dependence on interaction strength, can potentially be probed experimentally using MEGA2D \cite{tang2024chip}.
By adjusting the distance between a graphene sample on a substrate and another graphene sample on the Si pillar, one can potentially tune the interlayer interactions and directly observe these effects.

\subsection{Electronic Transition at Stronger Interlayer Interactions}

On the other hand, reducing the twist angle \cite{bennett2024twisted} or applying pressure \cite{carr2018pressure,yankowitz2018dynamic,yankowitz2019tuning} can enhance interlayer interactions and also modify the electronic properties.
We simulate the application of pressure by reducing the average layer separation of the relaxed structure of magic-angle tBLG by $\Delta z$, allowing us to study the effects of further decreasing the twist angle without performing calculations with even larger supercells \cite{carr2018pressure}.
Compressing the layers at the magic angle causes parts of the bands to approach each other, eventually touching and reopening at $\G$ and M (Fig.~\ref{Fig3}).
The closing and reopening of the energy separation at $\G$ is accompanied by an exchange of wavefunction character between the upper and lower flat bands throughout the mBZ. 
The probability density of the projection of the wavefunction onto layer 1 and sublattice $A$, $\Psi^{A_1}_{\Gfpm}$, shown below the band structure for each value of $\Delta z$ in Fig.~\ref{Fig3}, illustrates this exchange at $\Gfpm$.
While the total wavefunctions at $\Gfpm$ form ring-like structures around AA sites, the individual projections form structures with threefold symmetry, and the projected wavefunctions of the upper and lower bands have opposite orientations, denoted by cyan and purple borders.
When compression of the layers causes the bands to close and reopen, the orientation of the projected wavefunctions in each band reverses, indicating that their character has been exchanged.

To be more precise, we analyze the eigenvalues of the \textit{microscopic} sublattice-projected wavefunctions (i.e., $\Psi$ instead of $\vert\widetilde{\Psi}\vert^2$) under the valley-preserving 180$^\circ$ in-plane rotation about the $y$-axis, $\M_y\rv = \M_y(x,y,z) = (-x,y,-z)$, which we label as $\M_y$ to connect with Ref.~\cite{zou2018band} where it was shown to be a diagnostic of the chirality and fragile topology in twisted bilayer graphene.
The $\M_y$ symmetry, which has the effect of exchanging the layers and sublattices, is shown in the inset of Fig.~\ref{Fig4} and acts on the microscopic sublattice-projected wavefunctions as follows \cite{fang2019angle}:
\beq{eq:mirror-eigenvalue}
\M_y
\begin{pmatrix} 
\Psi^{A_1}(\rv) \\ 
\Psi^{B_1}(\rv) \\ 
\Psi^{A_2}(\rv) \\ 
\Psi^{B_2}(\rv) 
\end{pmatrix}
=
\begin{pmatrix} 
\Psi^{B_2}(\M_y \rv) \\ 
\Psi^{A_2}(\M_y \rv) \\ 
\Psi^{B_1}(\M_y \rv) \\ 
\Psi^{A_1}(\M_y \rv) 
\end{pmatrix} 
=
\la
\begin{pmatrix} 
\Psi^{A_1}(\rv) \\ 
\Psi^{B_1}(\rv) \\ 
\Psi^{A_2}(\rv) \\ 
\Psi^{B_2}(\rv)
\end{pmatrix}
\eec

where $\la = \pm 1$ is the $\M_y$ eigenvalue. Fig.~\ref{Fig4} shows projections of the microscopic wavefunctions at $\Gfpm$ before ($\D z = -0.075~\mathrm{\AA}$) and after ($\D z = -0.09~\mathrm{\AA}$) the closing and reopening of the energy separation.
Before the closing, the eigenvalues of the flat bands are $(\la_{\Gfm}, \la_{\Gfp}) = (+1,-1)$, and afterwards, they become $(-1,+1)$, demonstrating an exchange of symmetry in this transition.
In both cases, the eigenvalues in opposite flat bands have opposite sign, meaning the Dirac cones at K and K$'$ have the same chirality, as expected \cite{zou2018band}, and each eigenvalue changes sign after the transition.
We further note that the eigenvalues flip sign not only at $\G$ but at all $\kv$ that remain fixed under $\M_y$, namely the line along $\G$-- M. These results are identical for the two valleys. We validate our analysis using a continuum model based on Ref.~\cite{ezzi2024analytical}, shown in Fig.~S7 \cite{SM}.

Experimentally, this new transition should be accessible and tunable with pressure, which effectively scales the twist angle \cite{carr2018pressure}, making it possible to induce this transition reversibly in a single device. 
We estimate this transition to occur in magic angle tBLG at an applied pressure of 0.5--1 GPa (see Fig.~S5 (b) \cite{SM}) which can be achieved using a pressure cell \cite{yankowitz2018dynamic,yankowitz2019tuning} or MEGA2D \cite{tang2024chip}.
Since the wavefunction characters of the upper and lower flat bands exchange after the transition, we expect the effects of electron and hole doping to switch with the application of pressure, which should be measurable through transport \cite{yankowitz2018dynamic,yankowitz2019tuning} or microscopy measurements \cite{inbar2023quantum}.
For example, superconductivity in magic-angle tBLG is observed to be more robust for hole doping than electron doping \cite{balents2020superconductivity}.
However, for a twist angle of 0.93$^{\circ}$, below the angle at which the transition is expected to occur \cite{bennett2024twisted}, superconductivity is only observed with electron doping \cite{codecido2019correlated}, which may be correlated with an exchange of wavefunction character; additional work is needed to make an explicit connection. 
The exchange of $\M_y$ eigenvalues of the microscopic wavefunctions between the upper and lower flat bands potentially signals changes in the band topology, as changes in the parity of occupied electronic states are often related to changes in topological invariants \cite{fu2007topological,fang2012bulk,kruthoff2017topological}.
In future work, calculation of topological invariants (e.g., Chern number) using simpler models, such as a ten-band tight-binding model \cite{po2019faithful} fit to the \emph{ab initio} bands, can verify the topological nature of this transition.

\section{Discussion}
We have presented a detailed first-principles characterization of the low-energy wavefunctions of magic-angle tBLG on atomic and moiré scales.
Both the microscopic and moiré-scale wavefunctions provide valuable insight into important aspects of the low-energy physics of tBLG, including the dependence of the topology on interlayer interactions.
Starting from well separated, noninteracting graphene layers, we have demonstrated two distinct band inversions that occur as the interlayer separation decreases, effectively increasing the interlayer interaction.
The first occurs when the flat bands exchange character with the neighboring dispersive bands near $\G$: $\Gfm \leftrightarrow \Gdm$ and $\Gfp \leftrightarrow \Gdp$.
As the interlayer separation is decreased further, the flat bands invert among themselves: $\Gfm \leftrightarrow \Gfp$. 
Therefore, this second inversion further entangles bands originating above and below the Dirac point: $\Gfm \leftrightarrow \Gdp$ and $\Gfp \leftrightarrow \Gdm$.
Thus, minimal phenomenological models must include complementary bands above \textit{and} below the flat bands to accurately capture the symmetry of the electronic bands of tBLG.

In summary, our first-principles approach provides accurate, atomic-scale, and computationally viable simulations of the microscopic wavefunctions of tBLG with over 10,000 atoms per supercell.
This approach will serve as a powerful tool for characterizing the correlated phases observed in STM measurements of the wavefunctions \cite{nuckolls2023quantum}, such as intervalley coherent order and Kekulé patterns.

While our results provide a high level of detail of the structure and evolution of the flat-band wavefunctions in undoped tBLG, we note that our calculations are based on DFT, which is a single-particle approach.
As such, they do not capture strong correlation effects, which play an important role in tBLG.
Extending our approach to include many-body effects, finite doping, and the direct calculation of topological invariants are important directions for future work.\\

\section{Acknowledgments}
The authors acknowledge support from the Simons Foundation award No.~896626 and the US Army Research Office (ARO) MURI project under Grant No.~W911NF-21-0147.
A.Z.~is supported by the U.S.~Department of Energy, Office of Science, Office of Advanced Scientific Computing Research, Department of Energy Computational Science Graduate Fellowship under Award Number DE-SC0025528.
D.B.~acknowledges support from the NTU Startup Grant (Award Number 025661-00003).
M.M.E.A.~acknowledges support from the Ministry of Education, Singapore (Research Centre of Excellence award to the Institute for Functional Intelligent Materials, I-FIM, project No. EDUNC-33-18-279-V12), the National Research Foundation, Singapore, under its AI Singapore Programme (AISG Award No: AISG3-RP-2022-028).
E.~M.~acknowledges the hospitality of the Harvard Physics Department, where part of this work was done. 
Computing resources were provided by the Harvard University FAS Division of Science Research Computing Group and the Texas Advanced Computing Center (TACC) at The University of Texas at Austin.

\end{document}